\documentclass[prd,superscriptaddress,onecolumn,showkeys]{revtex4}
\usepackage{eurosym}
\usepackage{amsfonts}
\usepackage{array}
\usepackage{amsthm}
\usepackage{bm}
\usepackage{palatino}
\usepackage{mathpazo}
\usepackage{amssymb}
\usepackage{eurosym}
\usepackage{amsmath}
\usepackage{epsfig}
\usepackage{graphics}
\usepackage{changes}
\usepackage{color}
\usepackage{graphicx}
\usepackage[colorlinks=true,
            linkcolor=blue,
           urlcolor=black,
           citecolor=black]{hyperref}

\def\be{\begin{equation}}
\def\ee{\end{equation}}
\def\beq{\begin{eqnarray}}
\def\eeq{\end{eqnarray}}

\def\bes{\begin{eqnarray}}
\def\ees{\end{eqnarray}}

\begin{document}

\title{\textbf{Stability Analysis of Charged Rotating Black Ring}}

\author{Riasat Ali}
\email{riasatyasin@gmail.com}
\affiliation{Department of Mathematics, Government College,
University Faisalabad Layyah Campus, Layyah-31200, Pakistan}

\author{Kazuharu Bamba}
\email{bamba@sss.fukushima-u.ac.jp}
\affiliation{Division of Human Support System, Faculty of Symbiotic Systems
Science, Fukushima University,
Fukushima 960-1296, Japan}

\author{Muhammad Asgher}
\email{m.asgher145@gmail.com}
\affiliation{Department of Mathematics, The Islamia
University of Bahawalpur, Bahawalpur-63100, Pakistan}

\author{M.F. Malik}
\email{fawaadmalik1992@gmail.com}
\affiliation{Department of Mathematics, Government College,
University Faisalabad Layyah Campus, Layyah-31200, Pakistan}

\author{Syed Asif Ali Shah}
\email{asifalishah695@gmail.com}
\affiliation{Department of Mathematics,The University of Lahore 1-Km Raiwind RD, Sultan Town Lahore, Pakistan}

\begin{abstract}
We study the electromagnetic field equation along with the WKB approximation.
The boson tunneling phenomenon from charged rotating black ring(CRBR)
is analyzed. It is examined that reserve radiation consistent with CRBR
can be computed in general by neglecting back reaction and
self-gravitational of the radiated boson particle. The calculated
temperature depends upon quantum gravity and CRBR geometry. We also
examine the corrected tunneling rate/probability of boson particles
by assuming charge as well as energy conservation laws
and the quantum gravity. Furthermore, we study the graphical
behavior of the temperature and check the stability and instability of CRBR.
\end{abstract}

\date{\today}
\maketitle
\section{Introduction}
Hawking studied the emission rate of radiation of all particles (photons, neutrinos,
gravitons, positrons, etc.) from black hole (BH) and also computed the temperature of the BH horizon.

Tunneling is the semiclassical method in which boson particles have
ability to cross the black ring and black hole(BH) horizon. Recent analysis
show growing interest in  Hawking temperature as a process of
tunneling from black ring. The key component for examining modified tunneling radiation
in the classical behavior of imaginary portion that includes
the boson particles from the black ring horizon.

The tunneling radiation of charged fermion have been observed in BHs with
Newman-Unti-Tamburino(NUT) in \cite{M1}. The authors
concluded that the tunneling radiation
depends on the electric charge, magnetic charge, acceleration, mass and rotation
parameters, as well as the NUT parameter of the BHs pair.
The analysis of the tunneling radiation from the global monopole Reissner 
Nordst\"{o}m-de Sitter BH in \cite{R1} and concluded that 
the corrected temperature depend on the global monopole.
Sharif and Javed analyzed \cite{R2} radiation
spectrum through tunneling fermions in the BHs horizon family.
They concluded that the tunneling radiation must be depend on the horizon of outgoing positive
particles and cosmological horizon of incoming particles.
Sharif and Javed \cite{R3} devoted to analyzing the thermodynamic properties of
BH with NUT, rotation and acceleration parameters and observed that the quantity of
thermodynamics, such as surface gravity and the area-entropy relationship.
The method of quantum tunneling of boson particles tunnel through the more generalized
BHs horizons by applying the Proca equation has been analyzed \cite{R4} and 
examined that the tunneling radiation is independent of the
types of particles at which particles through the Kerr horizon.

The application of the Hamilton-Jacobi method and the significance of field equation with
electromagnetic interactions has been studied \cite{R5}.
The massive boson in the electric and magnetic tunneling field and the temperature
of BHs surrounded by fluid has been measured.
The Hawking radiation from Myers Perry 5D BHs and
black ring as a semiclassical tunneling have been investigated \cite{R6}.
It has been analyzed that the
Lagrangian charged field equation for charged boson and the WKB approximation.
Javed et al analyzed \cite{R8} the radiation for kinds of
Banados-Teitelboim-Zanelli-like BHs for higher dimensional spaces and
studied that the rotation parameter effect on the tunneling radiation.

The GUP plays very significant role to study the gravity effects (GUP corrections). To assume the
GUP effect on Hawking radiation, Dirac, Klein-Gordan and Lagrangian
equations will be corrected by assuming gravity effects.
Many authors have analyzed tunneling approach for different type particles with various spin
such as fermions, scalar and bosons through the horizons of different black rings, wormholes and BHs, they also studied
their corresponding modified Hawking temperature.
The BHs are the main experimental scalar and vector fields for studying the effects of gravity and
a many literature on the thermodynamical properties of BHs to study the gravity effects under GUP.
The BH thermodynamics have been analyzed within the GUP \cite{1,8}.
The field equation in the WKB approximation and tunneling of
bosons from the BHs in gauged super-gravity has been analyzed \cite{9}.
The GUP effect on BHs radiation and the instable and stable have
been analyzed.

This paper is organized as follows.
In Sec. {\bf II}, it contains the CRBR metric information, analyzes the imaginary part of boson particle action along the classical forbidden
trajectory and finally compute the modified temperature. 
In Sec. {\bf III}, we analyzed the graphical behavior of temperature. In last section, final results and summery are made.

\section{Rotating Black Ring}

The Einstein field equations in 5D have been expressed \cite{a1} a stationary
flat regular solution and have an horizon outside of topology $S^{1}\times S^{2}$.
The BH solution is proposed to have a naked
singularity, and the black ring solution also approach the similar naked singularity. 
A supergravity solution depicts a $S^{2}\times S^{1}$ CRBR horizon and its properties in a
5D asymptotically flat metric have been studied \cite{a2}.
For the CRBR, a curious relationship
between the charge and the mass has been concluded.
The 5D black ring horizon \cite{a23} has topology $S^{2} \times S^{1}$ and
self-gravitational and its tension are exactly balanced by the rotation of the black ring.
The charge parameter increases the stability of black rings.
The asymptotically flat stationary black ring \cite{a3} is a boundary value solution to the 5D
Einstein vacuum equations.
The asymptotically flat 5D black ring
solutions for the Einstein
vacuum equations has been tested taking three commuting Killing vector fields.
The stationary axisymmetric nonlinear $\sigma$-model with self-gravitating in 5D metric
taking three commutating Killing vector fields has been studied \cite{a4}.
The regular rotating horizon of black ring have been generalized by two angular momenta and mass.
A dipole charged black ring solution contains conical naked singularity and turns out to be generically unbalanced
has been analyzed \cite{a5}.
A dipole black ring can be used as an independent parameter to uniquely specify a black lens minimal supergravity solution.

The CRBR radiation have been investigated for
the significance field and consider in physics of gravitational as well
as in advanced theoretical frameworks for example brane/string
dynamics. The CRBR radiation pulling attention of physicists which shows that CRBR is extremely excited states
of quantum and so can be required to be fully realized in
conditions of quantum gravity.
Otherwise, the CRBR radiation is describe to
establish the thermodynamics law in spacetimes effecting
CRBR consistent. Generally approaches for the CRBR radiation,
which takes spectrum of thermal only.

In this section, we analyzed the problem of Hawking temperature
from CRBR result of the Einstein-Maxwell-dilation gravity
5-dimensional space theory.
The first law of thermodynamics exists for BRs and the emission probability is
associated for the existance of Bekenstein Hawking BRs entropy
during the radiation can be analyzed in Ref. \cite{111}. This solution is agree
with the results of Hawking temperature in conditions
of tunneling for other case of BHs and it appears
to be general axiom for the tunneling method. So, the BRs
are new class of spacetime. It is analyzed
on their different axioms and particular solution of quantum
state using quantum tunneling method (the CRBR quantum state has been observed,
to especially leading order is already observed in
previous paper). The stability analyzes both form of
thermodynamical and dynamical views, also the BRs
are concerning vector field of further work.

The Einstein-Maxwell-dilaton 5D BR theory for Hawking radiation are analyzed
by using the quantum tunneling method. To study the
correct temperature, we assume and developed the quantum tunneling
approach to calculate the effects of quantum gravity and charge.
The effect of back reaction is not taken in this phenomenon. 
The neutral and dipole cases line element of a single electric charge black ring is written as \cite{a2}
\begin{eqnarray}
ds^2&&=-\frac{G(y)}{G(x)K^2(x, y)}[dt-C(\nu, \lambda)
R\frac{1+y}{G(y)}cosh^2\alpha d\phi]^2+\frac{R^2}{(x-y)^2}\nonumber\\
&&G(x)[-\frac{F(y)}{G(y)}d\phi^2-
\frac{1}{F(y)}dy^2+\frac{1}{F(x)}dx^2+\frac{F(x)}{G(x)}\psi^2],
\end{eqnarray}
where $C(\nu, \lambda)=\sqrt{\frac{1+\lambda}{1-\lambda}
\lambda(\lambda-\nu)}$,~~$K(x, y)=\lambda(x-y)\sin h^2\alpha F(x)$,\\
$G(\xi)=1+\lambda\xi$,~~$F(\xi)=(1-\xi^2)(1+\nu\xi)$,
parameters $\nu$ and $\lambda$ are dimensionless and take the
values in range $1>\lambda\geq\nu>0$, we do not take
the conical singularity at $x=1$, $\lambda$ and $\nu$
are related to each other, say $\lambda=\frac{2\nu}
{1+\nu^2}$ and $\alpha$ is a parameter representing as the
charge of electric. The coordinate $\phi$ and $\psi$ are two cycles
of BR and $x$ and $y$ the values range $1\geq x\geq-1$
and $1\geq y\geq -\infty$ and $R$ has the length of dimensional.
The explicit computation of the electric charged is
$Q=\frac{2M sin h 2\alpha}{3(1+\frac{4}{3}sinh^2\alpha)}$.
The mass of BR is $M=\frac{3\pi R^2\lambda}{4(1-\nu)}$
in \cite{RR1} and its angular momentum takes the form
$J=\frac{\pi R^3 p\sqrt{\lambda(\lambda-\nu)(1+\lambda)}}{2(1-\nu)^2}.$\\

In order to study the tunneling probability for boson particles through
the CRBR horizon $y_{+}$, we will assume the Lagrangian equation with
gravity and electromagnetic effects. The charged motion with spin-1
 fields is described out by the given Lagrangian
gravity equation with vector field $\chi_{\mu}$ \cite{7}
\begin{eqnarray}
&&\frac{1}{\sqrt{-g}}\partial_{\mu}(\sqrt{-g}\chi^{\nu\mu})+
\frac{m^{2}}{h^{2}}\chi^{\nu}+\frac{i}{h}e A_{\mu}\chi^{\nu\mu}+
\beta\hbar^{2}\partial_{0}\partial_{0}\partial_{0}(g^{00}\chi^{0\nu})\nonumber\\
&&-\beta\hbar^{2}\partial_{i}\partial_{i}\partial_{i}(g^{ii}\chi^{i\nu})=0\label{L},
\end{eqnarray}

where $g$ is determinant of coefficients matrix, $m$ is
particles mass and $\chi^{\mu\nu}$ is anti-symmetric tensor, i.e.,

\begin{eqnarray}
\chi_{\nu\mu}&=&(1-\beta\hbar^{2}\partial^{2}_{\nu})\partial{\nu}
\chi_{\mu}-(1-\beta\hbar^{2}\partial^{2}_{\mu})\partial{\mu}\chi_{\nu}
+(1-\beta\hbar^{2}\partial^{2}_{\nu})\frac{i}{h}eA_{\nu}\chi_{\mu}\nonumber\\
&-&(1-\beta\hbar^{2}\partial^{2}_{\mu})\frac{i}{h}eA_{\mu}\chi_{\nu}.\nonumber
\end{eqnarray}
Here, $A_\mu$ is denoted as the electromagnetic vector potential of the
CRBR and $e$ denotes the charge of the boson particle.
Here $\beta$ is quantum gravity parameter.
The $\chi$ components are calculated as

\begin{eqnarray}
&&\chi^{0}=\frac{V}{f}\chi_{0}-\frac{Z}{f}\chi_{1},
~~\chi^{1}=\frac{-Z}{f}\chi_{0}-\frac{U}{f}\chi_{1},\nonumber\\
&&\chi^{2}=\frac{1}{W}\chi_{2},~~\chi^{3}=\frac{1}{X}\chi_{3},
~~\chi^{4}=\frac{1}{Y}\chi_{4},~~
\chi^{01}=\frac{Z^2\chi_{10}+UV\chi_{01}}{f^2},\nonumber\\
&&\chi^{02}=\frac{V\chi_{02}-Z\chi_{12}}{Wf},~~
\chi^{03}=\frac{-Z\chi_{03}+U\chi_{13}}{Xf},~~
\chi^{04}=\frac{-Z^2\chi_{04}+U\chi_{14}}{Yf},\nonumber\\
&&\chi^{12}=-\frac{Z\chi_{02}+U\chi_{12}}{Wf},~~
\chi^{13}=\frac{-Z\chi_{03}+U\chi_{13}}{Xf},~~
\chi^{14}=\frac{-Z\chi_{04}+U\chi_{14}}{Yf},\nonumber\\
&&\chi^{23}=\frac{1}{WX}\chi_{23},~~\chi^{24}=\frac{1}{WY}\chi_{24},
~~\chi^{34}=\frac{1}{XY}\chi_{34}\nonumber
\end{eqnarray}
where $f=UV-Z^2$.
The WKB approximation in \cite{RR3} is given by
\begin{equation}
\chi_{\nu}=c_{\nu}\exp[\frac{i}{\hbar}\Theta_{0}(t, \phi, y, x, \psi)+
\Sigma \hbar^{n}\Theta_{n}(t, \phi, y, x, \psi)].\nonumber
\end{equation}
Here $\Theta_{0}$, $\Theta_{n}$ and $c_{\nu}$ are arbitrary functions and
constant.
By neglecting the higher order terms for $n=1, 2, 3,...$
and applying Eq. (\ref{L}), we obtain the set of field equations.
The electromagnetic vector potential of black ring is given by
\begin{equation}
A_{\mu}=A_{t}dt+A_{\phi}d\phi,
\end{equation}
where
\begin{equation}
A_{t}=\frac{\lambda(x-y)\sin h\alpha \cos h\alpha}{G(x)K(x, y)},~~
A_{\phi}=\frac{C(\nu, \lambda)R(1+y)\sin h\alpha
\cos h\alpha}{G(x)K(x, y)}.\nonumber
\end{equation}
Using separation of variables technique \cite{R2}, we can choose
\begin{equation}
\Theta_{0}=-E_{0}t+J\phi+I(x, y)+L\psi,
\end{equation}
where $E_{0}=(E-\sum_{i=1}^{2}j_{i}{\Omega_{i}})$, $E$ is the energy
of particle, $J$ and $L$ are represent the particles angular
momentums corresponding to the angles $\phi$ and $\psi$ respectively.
From set of field equations can be  obtain a $5\times5$ matrix
equation $G(c_{0},c_{1},c_{2}, c_{3}, c_{4})^T =0,$ the components of the expected
matrix have the following form
\begin{equation*}
G(c_{0},c_{1},c_{2},c_{3},c_{4})^{T}=0,
\end{equation*}
which implies a $5\times5$ matrix denoted as "$G$", whose entries
are devoted as follows:
\begin{eqnarray}
G_{00}&=&\frac{Z^2}{f}[E_{0}^2
+\beta E_{0}^4]
-\frac{UV}{f}[J^{2}+\beta J^{4}]-\frac{V}{Wf}[I_{1}^{2}+\beta I_{1}^{4}]\nonumber\\
&&-\frac{V}{Y}[L^{2}+\beta L^{4}]-m^2 V,\nonumber\\
G_{01}&=&\frac{Z^2}{f}\tilde{A}J
-\frac{UV}{f}\tilde{A}J+\frac{Z}{W}[I_{1}^{2}+\beta I_{1}^{4}]
+\frac{Z}{Y}[L^{2}+\beta L^{4}]
-m^2 Z,\nonumber\\
G_{02}&=&-\frac{V}{W}\tilde{A}I_{1}-
\frac{Z}{Wf}[J+\beta J^{3}],~~
G_{03}=-\frac{V}{X}\tilde{A}I_{2}-
\frac{Z}{X}[J+\beta J^{3}],\nonumber\\
G_{04}&=&-\frac{V}{Y}\tilde{A}L, \nonumber\\
G_{10}&=&\frac{Z^2}{f}[E_{0}L
+\beta E_{0}J^3]
-\frac{UV}{f}[E_{0}J+\beta E_{0}J^{3}]+\frac{Z}{Wf}[I_{1}^{2}+\beta I_{1}^{4}]+\nonumber\\
&&\frac{Z}{X}[E_{0}+
\beta E_{0}]I_{2}
+\frac{eA_{0}Z^2}{f}[E_{0}+\beta E_{0}^{3}]
+{eA_{0}UV}[J+\beta J^{3}]+m^2 Z,\nonumber\\
G_{11}&=&\frac{Z^2}{f}E_{0}\tilde{A}
-\frac{UV}{f}E_{0}\tilde{A}-\frac{U}{W}
[I_{1}^{2}+\beta I_{1}^{4}]
-\frac{U}{X}[I_{1}^{2}+\beta I_{1}^{4}]I_{2}
-\frac{U}{Y}[L^{2}+\beta L^{4}]\nonumber\\
&&-m^2 ZX-eA_{0}Z^2\tilde{A}+{eA_{0}UV}\tilde{A},\nonumber\\
G_{12}&=&-\frac{Z}{W}\tilde{A}I_{1}
+\frac{U}{Wf}[J+\beta J^{3}]I_{1}+\frac{U}{X}[J+\beta J^{3}]I_{2},\nonumber\\
G_{13}&=&\frac{Z}{X}\tilde{A}I_{2},~~
G_{14}=\frac{Z}{Y}\tilde{A}L+
\frac{U}{Yf}[J+\beta J^{3}]L, \nonumber\\
G_{20}&=&-V[E_{0}I_{1}+
\beta E_{0}I_{1}^{3}]+\frac{Z}{f}[I_{1}+\beta I_{1}^{3}]J+
\frac{eA_{0}V}{Wf}[I_{1}+\beta I_{1}^{3}],\nonumber\\
G_{21}&=&\frac{Z}{f}[E_{0}
I_{1}+\beta E_{0}I_{1}^{3}]+\frac{Z}{f}[I_{1}+\beta I_{1}^{3}]J+
\frac{eA_{0}V}{Wf}[I_{1}+\beta I_{1}^{3}],\nonumber\\
G_{22}&=&-\frac{V}{f^2}E_{0}\tilde{A}
-\frac{Z}{f^2}E_{0}\tilde{A}-\frac{Z}{f}[E_{0} J+
\beta E_{0}J^{3}]-\frac{Z}
{f}\tilde{A}J-\frac{U}{f}[J+\beta J^{3}]\nonumber\\
&&-m^2-\frac{1}{X}[I_{2}^{2}+
\beta I_{2}^{4}]-\frac{1}{Y}[L^{2}+\beta L^{4}]-\frac{eA_{0}V}{f^2}\tilde{A}
+\frac{eA_{0}V}{f^2}[J_{1}+\beta J_{1}^{3}],\nonumber\\
G_{23}&=&\frac{1}{X}[I_{1}+\beta I_{1}^{3}]I_{2},~~
G_{24}=\frac{1}{Y}[I_{1}+\beta I_{1}^{3}]L,\nonumber\\
G_{30}&=&-\frac{V}{f}[
I_{2}+\beta I_{2}^{3}]E_{0}-\frac{Z}{f}[I_{2}+\beta I_{2}^{3}]J+
\frac{eA_{0}V}{f}[I_{2}+\beta I_{2}^{3}],\nonumber\\
G_{31}&=&\frac{Z}{f}[
I_{2}+\beta I_{2}^{3}] E_{0}+\frac{U}{f}[I_{2}+\beta I_{2}^{3}]J-
\frac{Z eA_{0}}{f}[I_{2}+\beta I_{2}^{3}],\nonumber\\
G_{32}&=&\frac{1}{W}[I_{2}+\beta I_{2}^{3}]I_{1},\nonumber\\
G_{33}&=&-\frac{V}{f}E_{0}\tilde{A}-
\frac{Z}{f}\tilde{A}J-\frac{U}{f}[J^{2}+\beta J^{3}]
-\frac{1}{W}[I_{1}^{2}+\beta I_{1}^{4}]
-\frac{1}{Y}[L^{2}+\beta L^{4}]\nonumber\\
&&-\frac{m^2}{X}
+\frac{eA_{0}V}{Wf}\tilde{A}+\frac{ZeA_{0}}{f^2}
[J+\beta J^{3}],\nonumber\\
G_{34}&=&\frac{1}{Y}[I_{2}+\beta I_{2}^{3}]L,\nonumber\\
G_{40}&=&-\frac{V}{f}[L+\beta L^{3}]E_{0}
-\frac{V}{f}[L+\beta L^{3}]J+
\frac{e A_{0}V}{f}[L+\beta L^{3}],\nonumber
\end{eqnarray}
\begin{eqnarray}
G_{41}&=&\frac{Z}{f}[L+\beta L^{3}]E_{0}+\frac{U}{f}[L+\beta L^{3}]J-
\frac{e A_{0}V}{f}[L+\beta L^{3}],\nonumber\\
G_{42}&=&\frac{1}{W}[L+\beta L^{3}]I_{1},~~
G_{43}=\frac{1}{X}[L+\beta L^{3}]I_{2},\nonumber\\
G_{44}&=&-\frac{V}{f}E_{0}\tilde{A}-\frac{Z}{f}
[J+\beta J^{3}]E_{0}
-\frac{Z}{f}\tilde{A}J-
\frac{U}{f}[J^{2}+\beta J^{4}]
-\frac{1}{W}[I_{1}^{2}+\beta I_{1}^{4}]\nonumber\\
&&-\frac{1}{X}[I_{2}^{2}+\beta I_{2}^{4}]-\frac{m^2}{Y}
+\frac{eA_{0}V}{f}\tilde{A}+\frac{ZeA_{0}}{f^2}[J+\beta J^{3}],\nonumber
\end{eqnarray}
where $\tilde{A}=E_{0}+
\beta E_{0}-eA_{0}-\beta eA_{0}E_{0}^2$, $J=\partial_{\phi}\Theta_{0}$, $I_{1}=
\partial_{x}{\Theta_{0}}$, $I_{2}=\partial_{y}{\Theta_{0}}$
and $L=\partial_{\psi}{\Theta_{0}}$.
For the non-trivial solution, the determinant $\textbf{G}$ is equal to
zero. So, we get
\begin{eqnarray}\label{a1}
ImI^{\pm}&=&\pm \int\sqrt{\frac{E_{0}^{2}
+X_{1}[1+\beta\frac{X_{2}}{X_{1}}]}{-\frac{f}{W}}}dy\\
&=&\pm i\pi\frac{E_{0}+[1+\Xi \beta]}{2\kappa(y_{+})},
\end{eqnarray}
where
\begin{eqnarray}
X_{1}&=&-\frac{2XZE_{0}}{f}J-
\frac{XZeA_{0}E_{0}^3}{f}J-\frac{XU}{f}J^2-\frac{X}{Y}L^{2}-m^2X,\nonumber\\
X_{2}&=&-\frac{XE_{0}^4}{f^2}-
\frac{XeA_{0}E_{0}^3}{f^2}-\frac{XZeA_{0}E_{0}^2}{f}J-
\frac{UX}{f}J^{4}-I_{2}^{4}\frac{XL^{{0}^{4}}}{Y}.\nonumber
\end{eqnarray}
The boson tunneling probability is given as
\begin{equation}
\Gamma=\frac{\Gamma_{emission}}{\Gamma_{absorption}}=
exp[{-4\pi}\frac{(E-\sum_{i=1}^{2}j_{i}\Omega_{i}-eA_{0})}
{2\kappa(y_{+})}][1+\Xi \beta].\label{Tu}
\end{equation}
We obtained the Hawking temperature at outer CRBR horizon, which is similar to the massless
particle. This method can also be applied
to the all other kinds of BRs.
The Hawking temperature CRBR is given as
\begin{equation}
T_{H}=\frac{\sqrt{(x-y)^2(1-x^2)(\nu x+1)\ell}}
{4\pi\sqrt{(1+y\nu)(1-y^2)}}[1+\Xi \beta]
\label{t}
\end{equation} where $\ell=\nu y^3+\nu y+\nu x-2xy-3xy^2 \nu+2$.
The CRBR temperature depends upon parameter
 $\nu$, $\beta$, $x$ and $y$. The modified temperature (if $\beta=0$) at which boson
particle tunnel through the CRBR horizon is similar to the temperature
of particles \cite{R6}.

\section{Gravitational Stability Analysis}

The positive boson particles are tunneling outside from the CRBR
horizon and there exist some fundamental interaction with them. The
negative boson particles are tunneling inside from the CRBR horizon.
The original spectrum of a boson will have the probability that inside the
surrounding or nearby region of a CRBR and negative particle must be
absorbed and boson particle would be irradiated and so many spectrum of
tunneling particles would be almost totally disseminated.
The condition of ionized (positive) boson particle is formed and these
particle will be radiated.

This section gives the analysis of corrected Hawking temperature
with rotation parameter for different values of gravity parameter.
The calculated Hawking temperature $T_{H}$
depends upon metric parameters ($\nu$, $x$, $y$) as well as on the
quantum gravity.

We also concluded that
$T_{H}$ does not depend upon the mass of the CRBR but only depends upon the quantum gravity
of the outgoing boson particles which is due to the influence of gravity.
The $T_{H}$ increases due to quantum gravity and remains constant when $0\leq y\leq 0.65$ and CRBR remains
stable in this range. The $T_{H}$ increases
when $0.65<y<\infty$ and CRBR is unstable in this range.

\begin{figure}\begin{center}
\epsfig{file=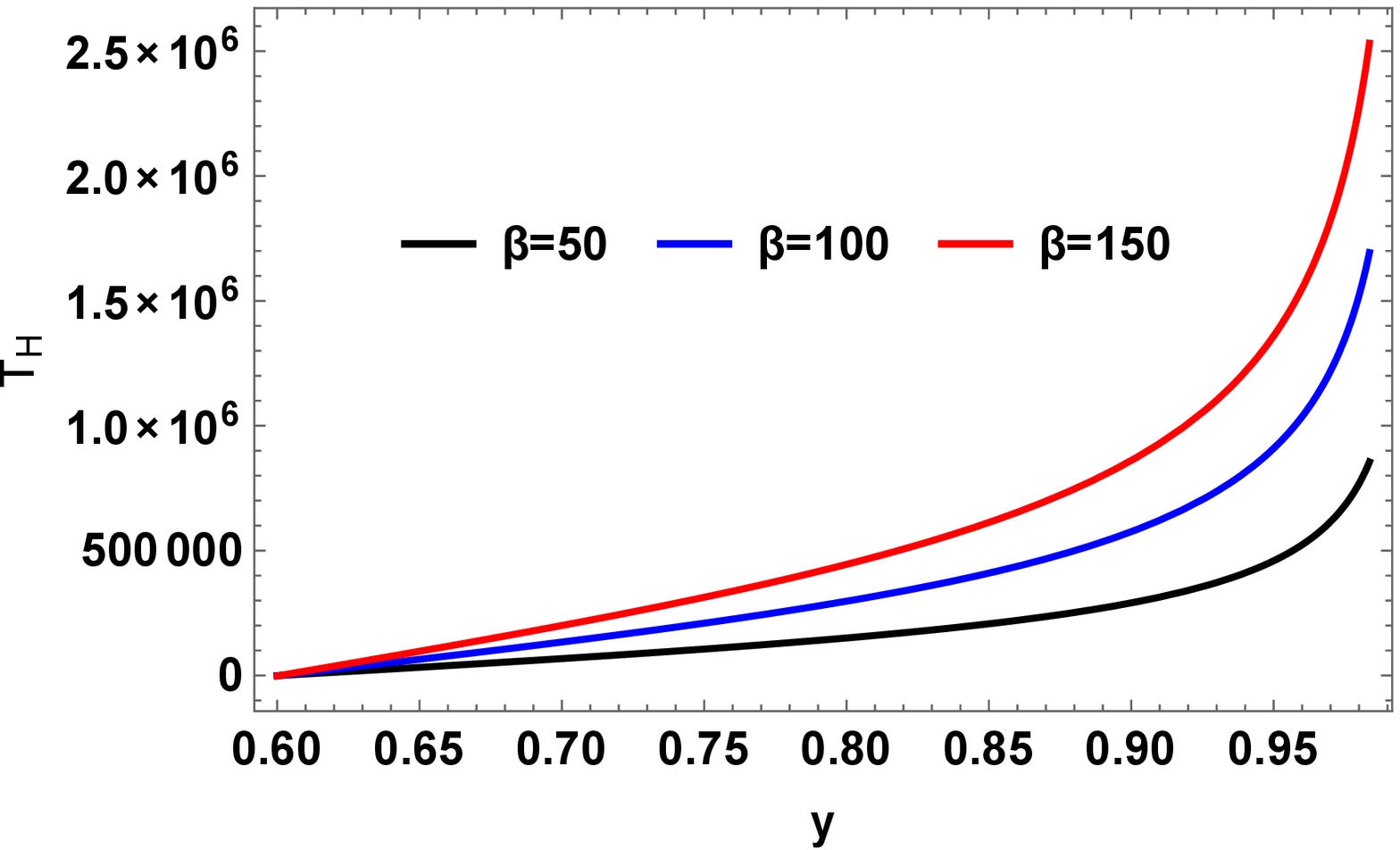, width=0.50\linewidth}
\caption{Hawking temperature $T_{H}$
versus $y$ for $x=2$, $\nu=10$, $\beta=50 ~to ~150$
and $\Xi=1$}.
\end{center}
\begin{center}
\epsfig{file=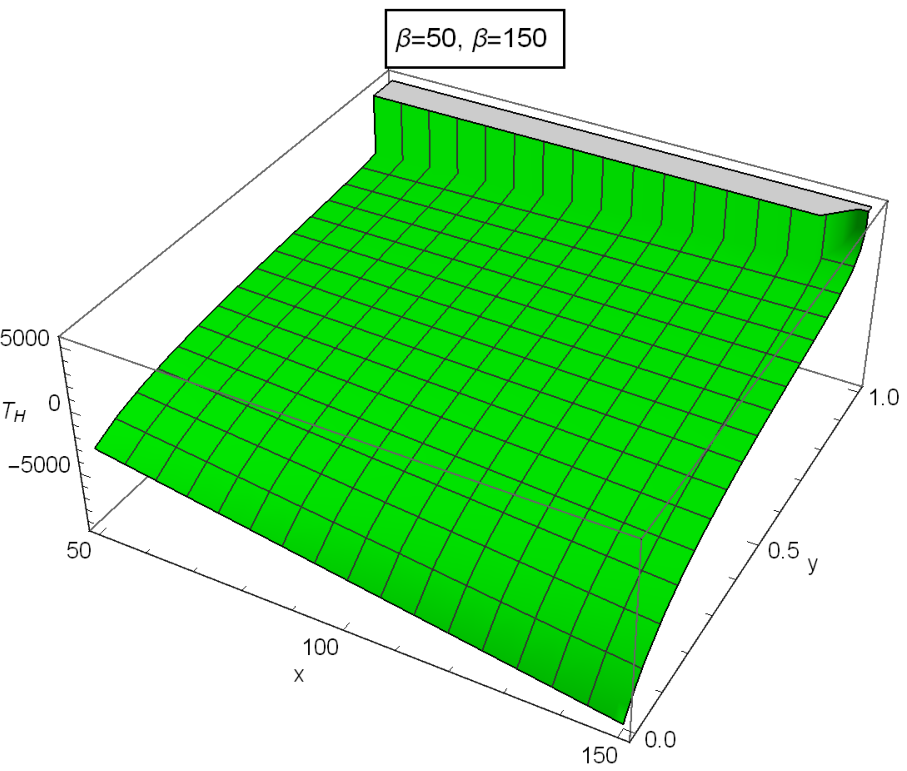, width=0.6\linewidth}
\caption{Hawking temperature $T_{H}$
versus $x$ and $y$ for $\nu=2$, $\beta=50 ~to ~150$
and $\Xi=1$}
\end{center}
\end{figure}
\section{Conclusions}

In this work, we have applied the Hamilton-Jacobi method to compute
tunneling probability of boson particles in the CRBR horizon.
Law of energy and angular momentum effects are conserved.
The back-reaction and self-interaction effect are
neglected. For the 5-dimensional CRBR, we computed that the
modified tunneling probability and modified temperature are
not only associated on the horizon of the 5-dimensional
CRBR but also the quantum gravity.
When $\beta\geq 50$ than the positive
modified temperature is received. In our analysis, we
obtained that the quantum corrections accelerate the high gain values in
modified temperature during the boson particles emission. Here, it is significant
to discuss that the value of the modified temperature is larger than
the original temperature and the CRBR stops emitting particles
when its mass approaches to the smaller value.
The Eq. (\ref{t}) proved that CRBR radiation is a like BH radiation. It shows that
the CRBR will radiate all types of particles as like BH radiation.
Our result gives a correction to the Hawking
temperature of CRBR. The result discovered out to the CRBR temperature
depends on the quantum gravity.

The results of corrected tunneling radiation in Eq. (\ref{Tu}) for boson
particle looks similar in \cite{R6} Eq. (2.14) if $\beta=0$,
but the mass, charge and angular momentum are same. Moreover, the CRBR
with the quantum gravity have more temperature than CRBR without quantum
gravity. We can observe that the larger quantum gravity supply gives more radiation. Moveover,
this solution is even satisfy if background CRBR metric is more generalized.
From our CRBR graphical analysis, the temperature increase with the increasing
CRBR horizon and CRBR reflects the stable condition
for small values of quantum gravity parameter.\\
{\bf Acknowledgments;}\\
The work of KB was partially supported by the JSPS KAKENHI Grant Number JP
25800136 and Competitive Research Funds for Fukushima University Faculty
(19RI017).\\

\end{document}